\documentclass[12pt]{article}
\usepackage{geometry} 
\usepackage{epstopdf}
\usepackage{graphics}
\usepackage{epsfig}
\usepackage{amsmath}
\usepackage{booktabs}
\usepackage{subfig}
\usepackage{natbib,hyperref}         
\title{Identifying Highly Correlated Stocks Using the Last Few Principal Components} 

\author{Libin Yang$^{1}$,   William Rea$^{1,*}$,  and Alethea Rea$^{2}$, \\
1. Department of Economics and Finance, University of Canterbury, \\
New Zealand \\
2. Data Analysis Australia, Perth, Australia \\
* Corresponding Author: email bill.rea@canterbury.ac.nz \\
 Phone +64-3-364-3474}

\begin{document}

\maketitle

\begin{abstract}
We show that the last few components in principal component analysis of
the correlation matrix of a group of stocks may contain useful financial
information by identifying highly correlated pairs or larger groups of
stocks. The results of this type of analysis can easily be included in
the information an investor uses to manage their portfolio.
\end{abstract}

\begin{description}
\item[Keywords: ] Principal component analysis, stock selection, 
diversification, stock portfolios
\item[JEL Codes: ] G11
\end{description}

\section{Introduction}

In recent years Principal Component Analysis (PCA) has been widely applied
to the study of financial markets. 
PCA \citep{Jolliffe1986} is a standard method in statistics
for extracting an ordered set of uncorrelated
sources of variation in a multivariate system. Given that financial
markets are typically characterised by a high degree of multicollinearity,
implying that there are only a few independent sources of information
in a market, PCA is an attractive method to apply.

Using random matrix theory, or, more
specifically, the spectral decomposition theorem \cite[][p13]{Jolliffe1986}
many authorities have divided the eigenvectors into three distinct
groups based on their eigenvalues. For example
\cite{kim2005} decomposed a correlation matrix of 135 stocks
which traded on the New York Stock Exchange (NYSE) into three parts:
\begin{enumerate}
\item The first principal component (PC1) with the largest 
eigenvalue which they asserted represents 
a market wide effect that influences all stocks.
\item A variable number of principal components (PCs) following the 
market component 
which represent synchronized fluctuations affecting 
groups of stocks.
\item The remaining PCs indicate randomness in the price fluctuations.
\end{enumerate}
Authorities such as \cite{Driesson2003}, \cite{kim2005}, \cite{C2007},
\cite{Kritzman2011}, \cite{Billio2012}, and \cite{Zheng2012} all
assumed
without any real question that the PCs of interest were PC1 and, depending
on the author's purposes, some or all of the PCs in group (2) above.

It is widely understood in the statistical literature that there may be 
two more categories of PC in addition to the three listed above. 
The first is the detection of outlier variables (rather than outlier
observations). In the analysis of our data below we did not find any
PCs of this type so will not pursue this further, see 
\citet[][Ch10]{Jolliffe1986} for details on this type of PC.

The other type of PC is near constant relationships between variables. 
To see how
these are detected, consider two stocks which are highly correlated.
Assume the eigenvalue of principal component $k$ is small and very close 
to zero. 
The eigenvalue of each principal component is a linear combination of all 
variables \citep{Jolliffe1986}, which can be written as
\begin{equation}
\alpha_{k}'\textbf{x} = \sum_{i =1}^{p} \alpha_{ki}x_{i}
\label{eqn:eigenvalue}
\end{equation}
where $\alpha_{k}'\textbf{x} $ is the eigenvalue of 
component $k$, and $\alpha_{ki}$ is the coefficient of variable
$i$ (in our case stocks) in component $k$.
If variables $x_{1}$ and $x_{2}$ are the two highly correlated 
variables (stocks) being detected in component $k$, each variable will have a 
large coefficient while the remainder of the variables will have
near zero coefficients. Equation \eqref{eqn:eigenvalue} then reduces
to
\begin{equation}
0 \approx \alpha_{k1}x_{1} + \alpha_{k2}x_{2} + 0.
\end{equation}
As a consequence, the closer $\alpha_{k1}$ and $\alpha_{k2}$ are in 
magnitude, the more correlated the $x_{1}$ and $x_{2}$.
If $x_1$ and $x_2$ are highly positively correlated then   
$\alpha_{k1}$ and $\alpha_{k2}$  will have opposite signs. 
If they are negatively correlated they will have the same sign. 
In this illustration we have used two stocks but
larger associations may be found.

In any given market such correlated assets may or may not exist.
If they do exist, finding them is straight-forward as we will show below and
the implications for stock selection and portfolio management are easily
understood.

The remainder of this paper is structured as follows; Section 
(\ref{Sec:DataDescript}) describes the data and methods, Section
(\ref{sec:Results}) presents our results and Section (\ref{sec:Conclude})
concludes.

\section{Data and Methods}\label{Sec:DataDescript} 

\subsection{Data}
Our research is based on the Australian market. The main index
for the market is the ASX200, which
is a market capitalization weighted index of the 200 largest 
shares by capitalization listed on the Australian Securities Exchange. 
The index, in its current form, was created on 31 March 2000. 
We investigated the 
constituents of the ASX200 index from inception to February 2014. 
The ASX200 index is a capitalization index and so
does not adjust for dividends.  In our
research we calculated the returns for
all constituents which included the dividends paid.

There was a high frequency of stocks that were added to or deleted from the 
index over time, so we identified all stocks which had been in the 
ASX200 for the whole study period. After adjusting for mergers, acquisitions, 
and name changes we obtained a final data set of 524 unique stocks. 
We obtained daily closing prices and dividends for each stock from the 
SIRCA database\footnote{\url{http://www.sirca.org.au/}}. 
All the prices and dividends were adjusted to be based on the AUD. 
The return was calculated in the following steps:
\begin{enumerate}
\item We created a new variable associated with each stock
called the Dividend Factor. 
We started with a factor of 1 and every time a dividend was paid we 
multiplied the Dividend Factor, 

\begin{align*}
\text{Daily Dividend Factor}_{i}(t) & = \left\{ 
  \begin{array}{l l}
    1 & \quad \textrm{if no dividend}\\
    1+\frac{D_{i}(t)}{P_{i}(t)} & \quad \text{if dividend}
  \end{array} \right\} \\
\text{Cumulative Dividend Factor}_{i}(t)&= \prod_{j=1}^t (\text{Daily Dividend Factor}_{i}(t))
\end{align*}

where $D_{i}(t)$ is the dividend for stock $i$ in time $t$, $P_{i}(t)$ is 
price for stock $i$ at time $t$ in units of one trading day.
\item We adjusted the price series with the dividend factor, the adjusted 
price was calculated by
$$
\text{PNEW}_{i}(t)= P_{i}(t) \times \text{Cumulative Dividend Factor}_{i}(t).  
$$
\item The return series for a given stock $i$ was calculated as
\begin{equation}
 \text R_{i}(t)=\frac{\text{PNEW}_{i}(t+1)-\text{PNEW}_{i}(t)}{\text{PNEW}_{i}(t)}.
\label{eqn:returns}
\end{equation}
\end{enumerate}

We extracted a set of stocks that had complete return information for the 
whole study period, and there were 156 such stocks. The remaining 368 
stocks were either listed after April 2000 or delisted before February 2014. 

\subsection{Principal Component Analysis}\label{sec:PComponents}

PCA can be applied to either a correlation matrix or a covariance matrix.
All PCAs reported in this paper were carried out on correlation matrices
generated from the return series.

Correlation matrices were generated with the {\tt cor} function in
the \verb+stats+ package,
PCAs were carried out using the function 
{\tt eigen} in base
{\tt R} \citep{R}.

We made biplots of the last few PCs using plotting functions in
the \verb+graphics+ package in base \verb+R+ 
and examined them for near constant relationships. See 
\citet[][Sec.\ 5.3]{Jolliffe1986} for details on biplots.

\section{Results}\label{sec:Results}

In this section, we present some details on the eigenvalues of the
last six PCs: bi-plots of their loadings,
which successfully picked up groups of 
stocks with highly correlated returns, together with time series plots
of their adjusted priced. We start with Principal Components 151 and 152
which picked up three pairs of near linear relationships and then discuss the 
``big four'' banks and two mining firms.

These six low variance principal components  detected stocks 
with high correlations.  In some applications the eigenvalues associated with 
the last few principal components are very
close to zero.  In our case the eigenvalues  the last few principal
components were small but clearly different from
zero. Nevertheless, they still picked up near linear relationships between
some stocks, see Table (\ref{table:eigenlast}).

\begin{figure}
\caption{Bi-plots of relative weights of each stock in components 151  and
152 arising from a PCA on a correlation matrix from the whole study period.
The stocks are colour coded using the Industry Classification Benchmark
Industry (ICB) classification.
Financials are Blue (33 stocks), Health Care are Red (9 stocks), Industrials
are Yellow (24 stocks), Consumer Services are Brown (19 stocks),
Basic Materials are Green (31 stocks), Oil\&Gas are Purple (16 stocks),
Utilities are orange (5 stocks), Consumer Goods are Black (9 stocks),
Telecommunications are Orchid (4 stocks), Technology are Grey (6 stocks).
Stocks with a loading of at least 0.2 in one of the PCs are labelled
with their ticker symbol.}
\label{fig:Components151-152}
{\includegraphics[width=\linewidth]{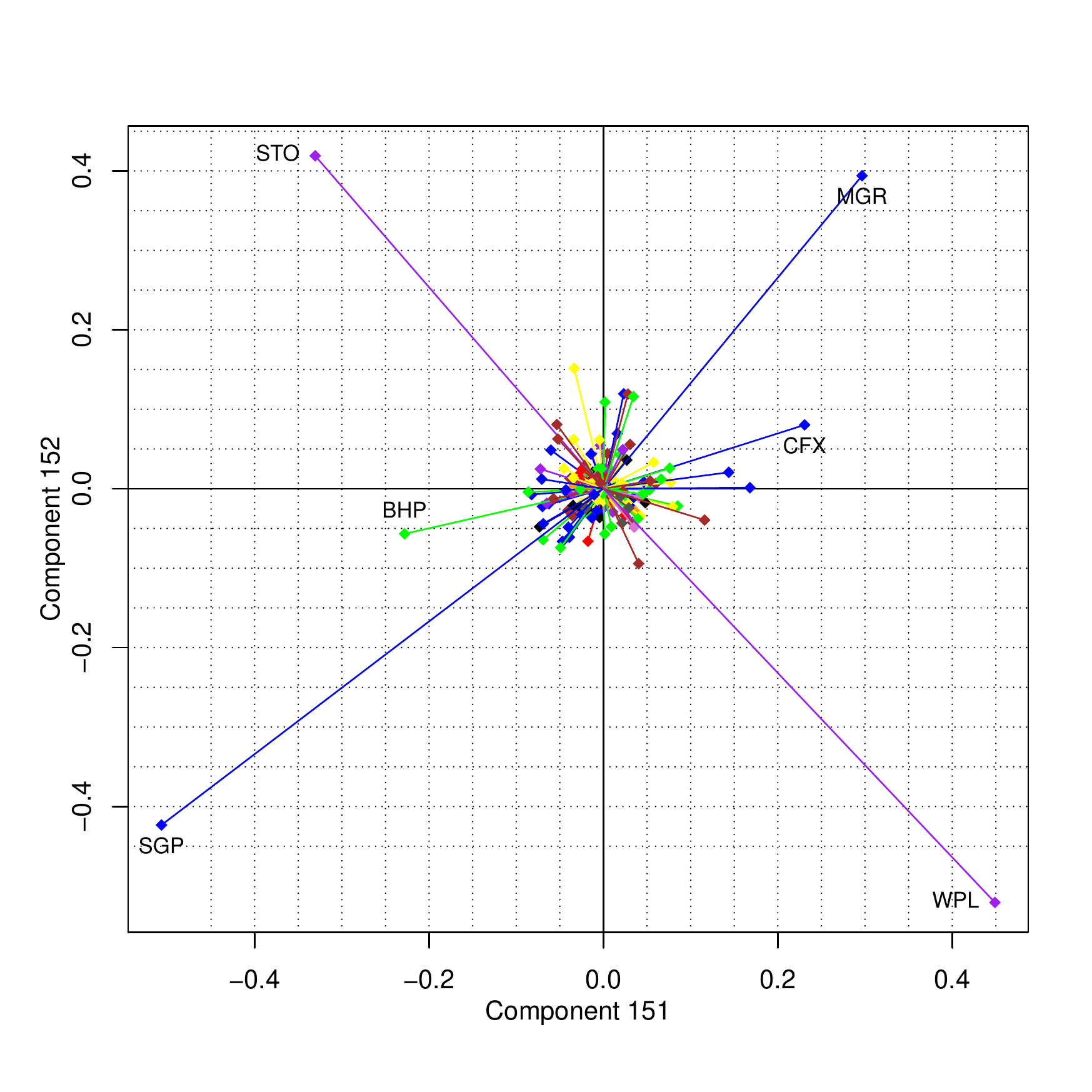}}
\end{figure}

In Figure (\ref{fig:Components151-152}), we present
biplots of PCs 151 and 152. BHP Billiton 
(BHP) and CFS Retail Property
Trust Group (CFX) in the real estate industry (CFX changed its name to Novion
Property Group after the close of the study period and now has the ticker symbol
NVN) clearly form a pair, they have high loadings of opposite signs
on PC151 but low loadings on PC152.
Mirvac Group  (MGR), Stockland (SGP), Santos Limited (STO) and Woodside
Petroleum Limited (WPL)  form a group of four 
and have high loadings on both PC151 and PC152. This
group of four can be broken into two pairs, STO with WPL and MGR with
SGP based on the signs of their loadings in PC151 and PC152.  


Curiously the first pair of stocks are not in the same industry.
BHP is in Basic Materials and CFX in the real estate industry. 
They tended to move in the same direction from the start of the
study period until 2011. Their price trajectories then began to move
in different directions after this time, see Figure (\ref{fig:CFXBHP}). 

The second pair, MGR and SGP, are both
large diversified real estate groups. 
In the beginning of 2000, they had nearly the same stock 
price level and have diverged since then. The similarity in their
price movements can be easily seen over short time frames,
but over the longer term 
their price level has diverged, see  Figure (\ref{fig:MGRSGP}). 

\begin{figure}
\caption{Time series plot of two stocks in Oil \& Gas industry identified
in the biplot of components 151 and 152: Santos (STO)
 and Woodside Petroleum (WPL).} \label{fig:WPLSTO}
\centering
{\includegraphics[width=\linewidth]{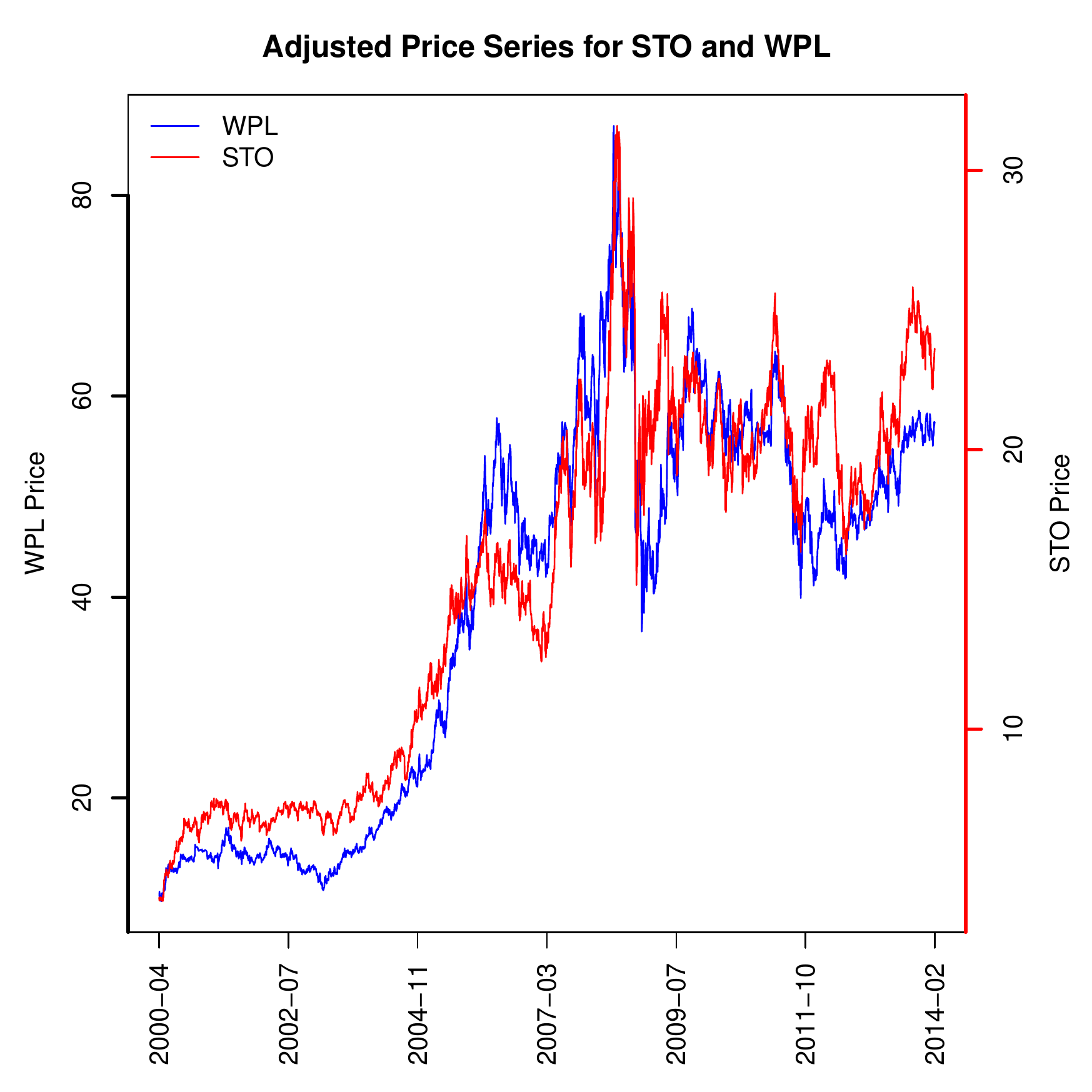}}
\end{figure}

The third pair of stocks are STO and 
WPL, which are in the Oil \& Gas industry. Both 
companies explore for and produce oil and gas from onshore and 
offshore wells.  The high correlations in their price
movements over both the short and long term are clearly evident,
see Figure (\ref{fig:WPLSTO}).

\begin{figure}
\caption{Bi-plots of relative weights of each stock in components 153  and
154 arising from a PCA on a correlation matrix from the whole study period.
The stocks are colour coded using the colour scheme described in
Figure (\ref{fig:Components151-152}).
Stocks with a loading of at least 0.25 in one of the PCs are labelled
with their ticker symbol.}
\label{fig:Components153-154}
\centering
{\includegraphics[width=\linewidth]{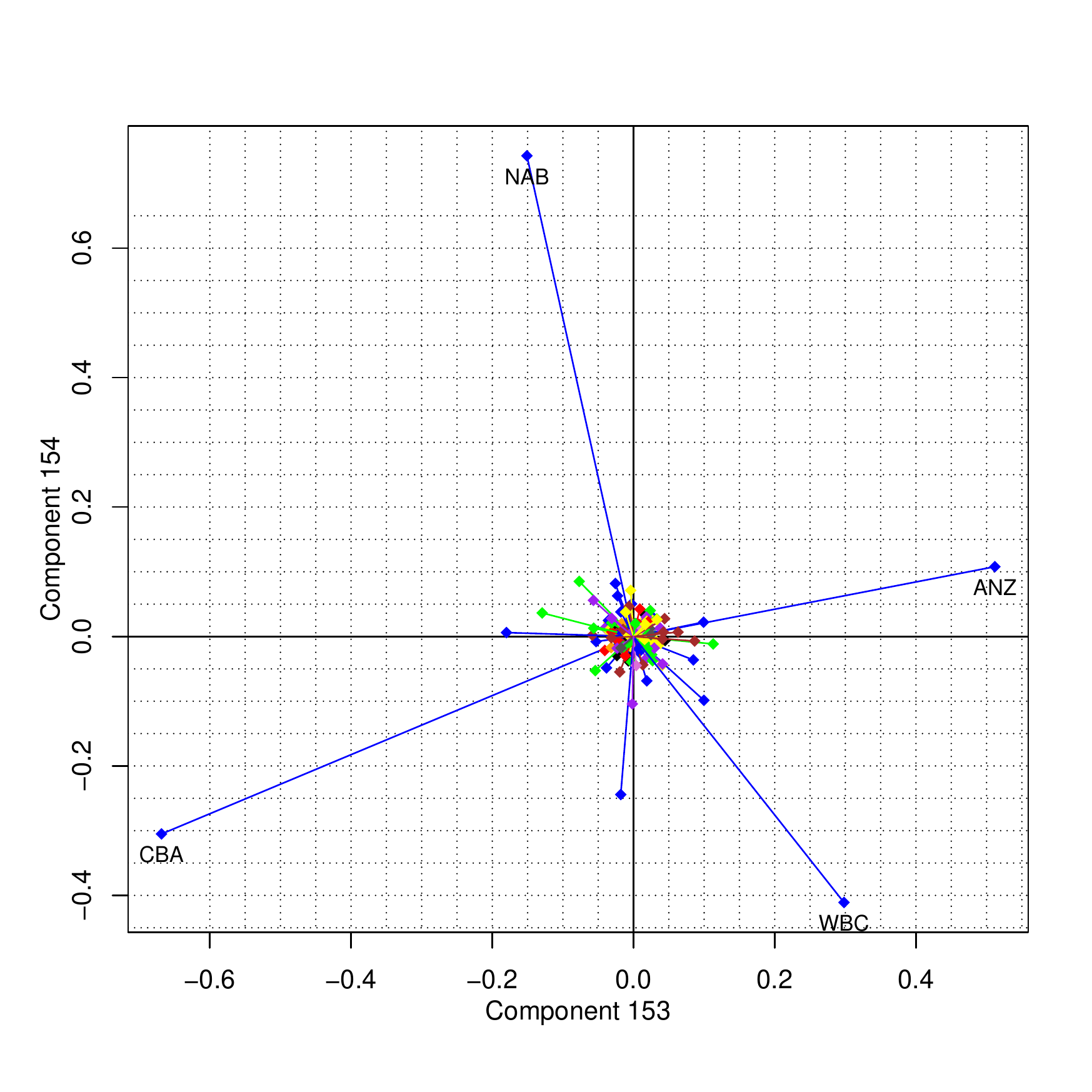}}
\end{figure}

\begin{figure}
\caption{Bi-plots of relative weights of each stock in components 155  and
156 arising from a PCA on a correlation matrix from the whole study period.
The stocks are colour coded using the colour scheme described in
Figure (\ref{fig:Components151-152}).
Stocks with a loading of at least 0.2 in one of the PCs are labelled
with their ticker symbol.}
\label{fig:Components155-156}
{\includegraphics[width=\linewidth]{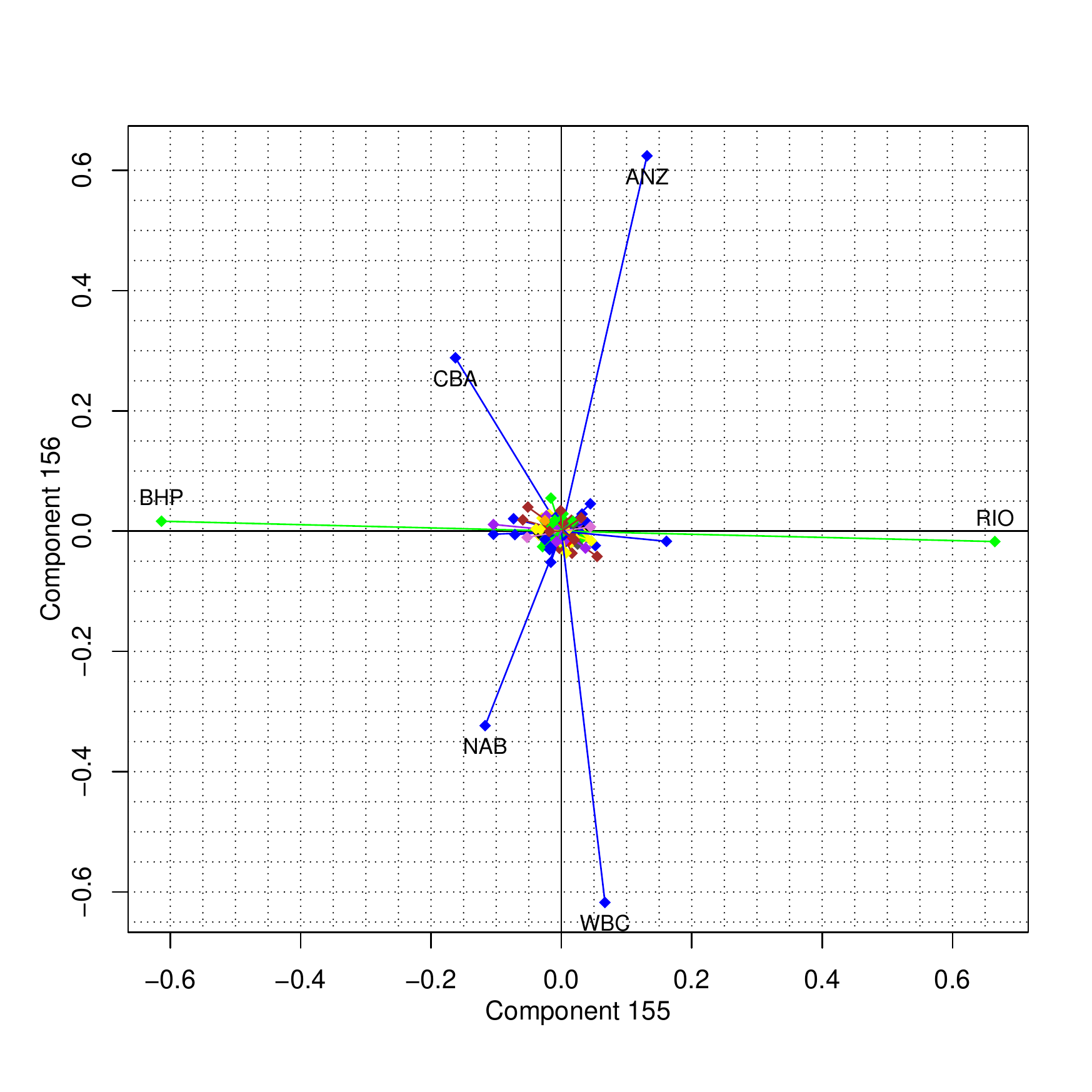}}
\end{figure}

The last four components (PC 153-156, Figures \ref{fig:Components153-154}
and \ref{fig:Components155-156}) all picked up the four largest banks 
in Australia, often referred to as the ``four pillars'',
their ticker symbols are
ANZ, CBA, NAB and WBC. 
The strong  relationships in price (and consequently returns) are easily seen
in Figure (\ref{fig:fourbigbanks}). To help visualizing the price co-movement 
of the four big banks, we used a different scale for CBA. Its 
dividend-adjusted
price 
changed from \$20 in the beginning of our study period to approximately 
\$150 at the end of study period while the other three banks had price 
levels that ranged from \$10 to \$70. Obviously, NAB was least correlated 
with others among the four banks. But after the 2008 financial crisis, 
all four banks converged to move very similarly.

\begin{figure}
\centering
\caption{ Time series plot of two stocks in Basic Materials identified in
component 155 : BHP and Rio Tinto (RIO). }
\label{fig:BHPandRIO}
\centering
\includegraphics[width=\linewidth]{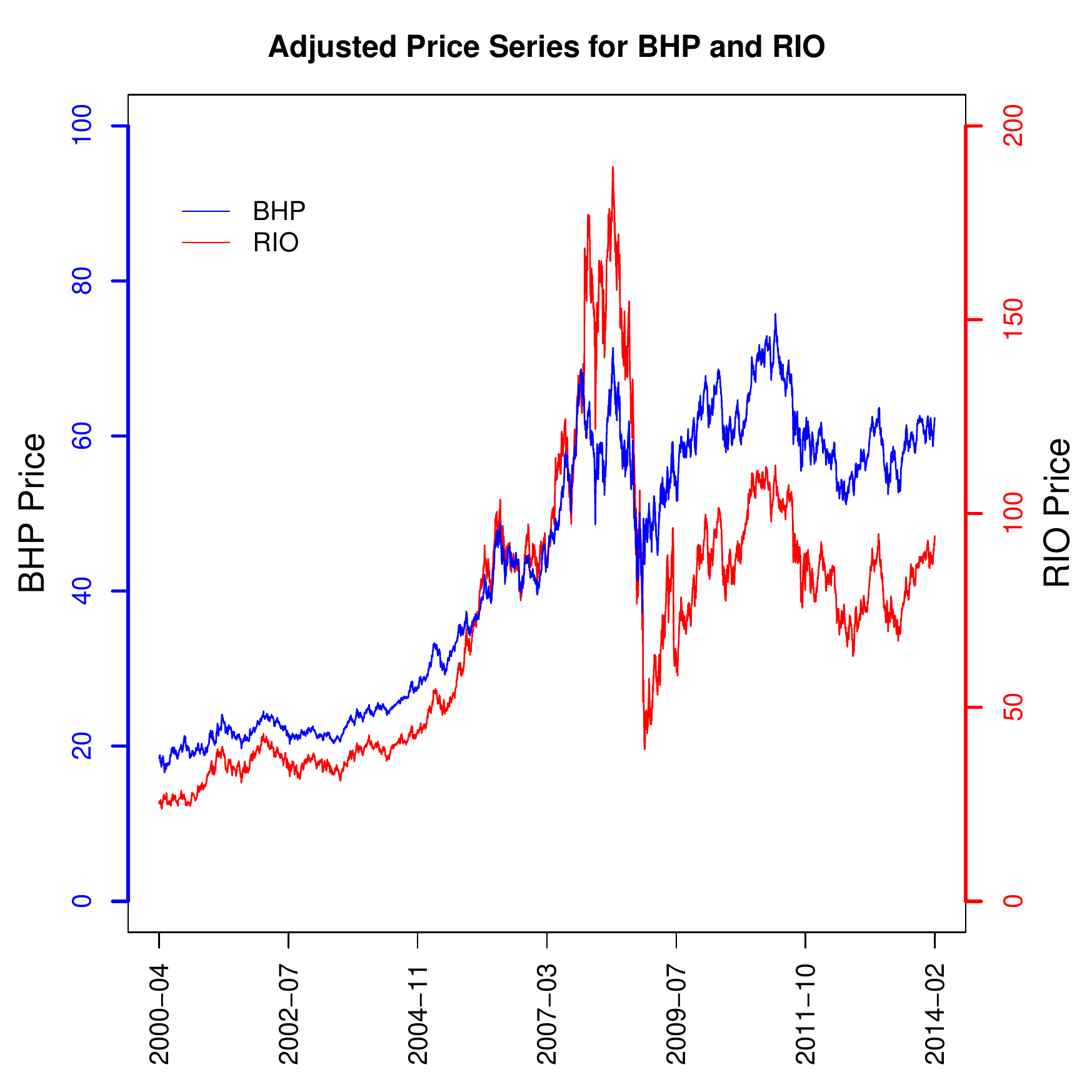}
\label{fig:BHPRIO}
\end{figure}

In the bi-plot of PC155 and PC156, Figure (\ref{fig:Components155-156}), 
Australia's two biggest 
mining firms were also picked up, BHP and Rio Tinto 
Ltd (RIO). However, they are clearly different from the four banks because
they have high loadings of opposite signs in PC155 and near zero loadings
in PC156. A plot of their price trajectories is presented
in  Figure (\ref{fig:BHPRIO}). 
At the beginning of our study 
period, the price of RIO was approximately 1.4 times of BHP. Before the 
price collapsed in 2008, both two stocks increased significantly and RIO 
increased even more. At the end of 2007, the price of RIO was about 
2.5 times of BHP. However, during the 2008 financial crisis, RIO also 
declined more than BHP. At the end of our study period, the price of 
RIO was about 1.5 times of BHP, which is almost the same as it was at 
the beginning of our study period.

\section{Conclusions}\label{sec:Conclude}

Our results above differ from that of other authorities such as
\cite{kim2005} who reported that only the market component (PC1)
and the subsequent
group PCs contained useful information about the financial market analysed. 
Our results illustrate that further financially useful information may be
contained in
the last few principal components because these may identify stocks 
with near linear correlations. This fact 
seems to have been overlooked in the finance literature despite it
being well known in the the statistics literature. 

The identification of highly correlated stocks (or other assets) can aid 
the task of portfolio selection and management because it identifies
pairs or groups of stocks which provide little benefit for diversification;
holding one of the pair or group will provide most of the benefits of
diversification while freeing funds to be invested in other assets. It is
not a given that such associations exist in any particular market. But if
they do exist they can be easily detected and that information fed into
the fund manager's or other investor's task of managing their portfolio
well.

\begin{figure}
\caption{Time series plot of two stocks identified in the biplot of
components 151 and 152: BHP-Billiton
(BHP) in basic materials and CFS Retail Property
Trust Group (CFX) in the real estate industry. }
\label{fig:CFXBHP}
\centering
\includegraphics[width=\linewidth]{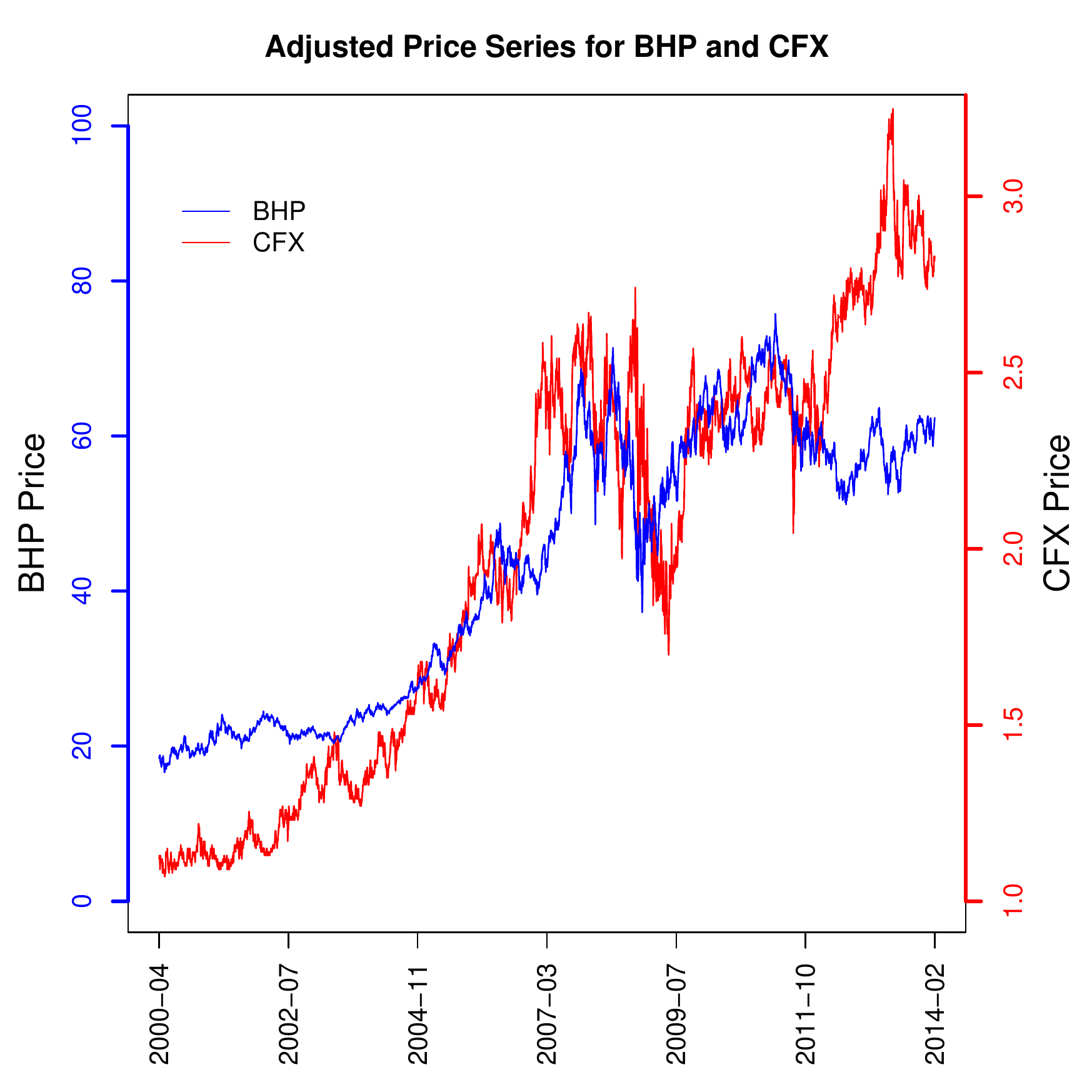}
\label{fig:stockpriceplot1}
\end{figure}

\begin{figure}
 \caption {Time series plots of near linear correlated stocks identified
in a biplot of components 151 and 152: Mirvac Group (MGP) and Stockland (SGP),
two stocks in the real estate sector.}
\centering
\centering
{\includegraphics[width=\linewidth]{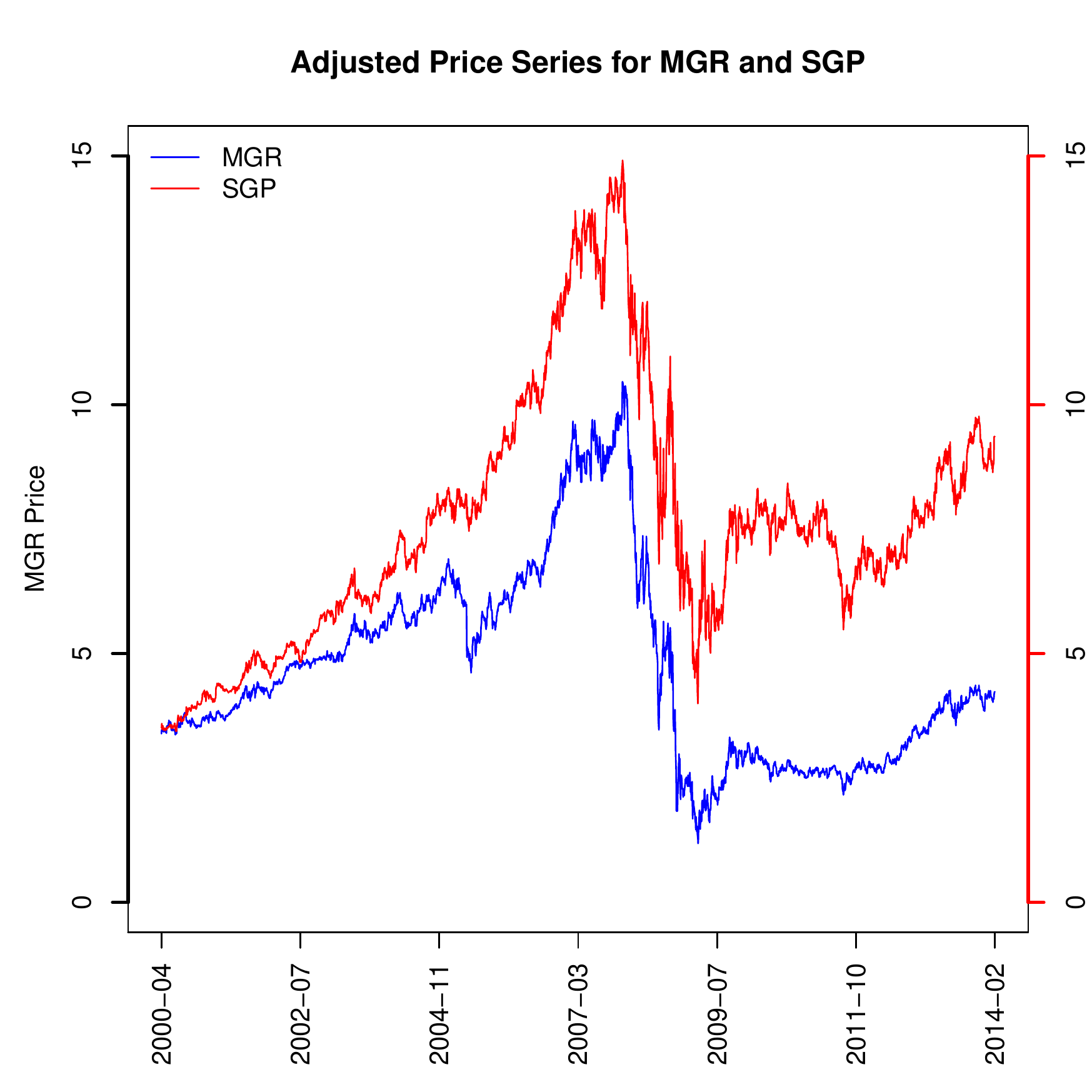}}
\label{fig:MGRSGP}
\end{figure}

\begin{figure}
\centering
\caption{Time series plots of near linear correlated stocks identified in
the biplots of
components 153 to 156: the four big banks in Australia. ANZ (ANZ), Commonwealth
Bank of Australia (CBA), National Australia Bank (NAB) and Westpac (WBC). }
\label{fig:fourbigbanks}
\centering
{\includegraphics[width=\linewidth]{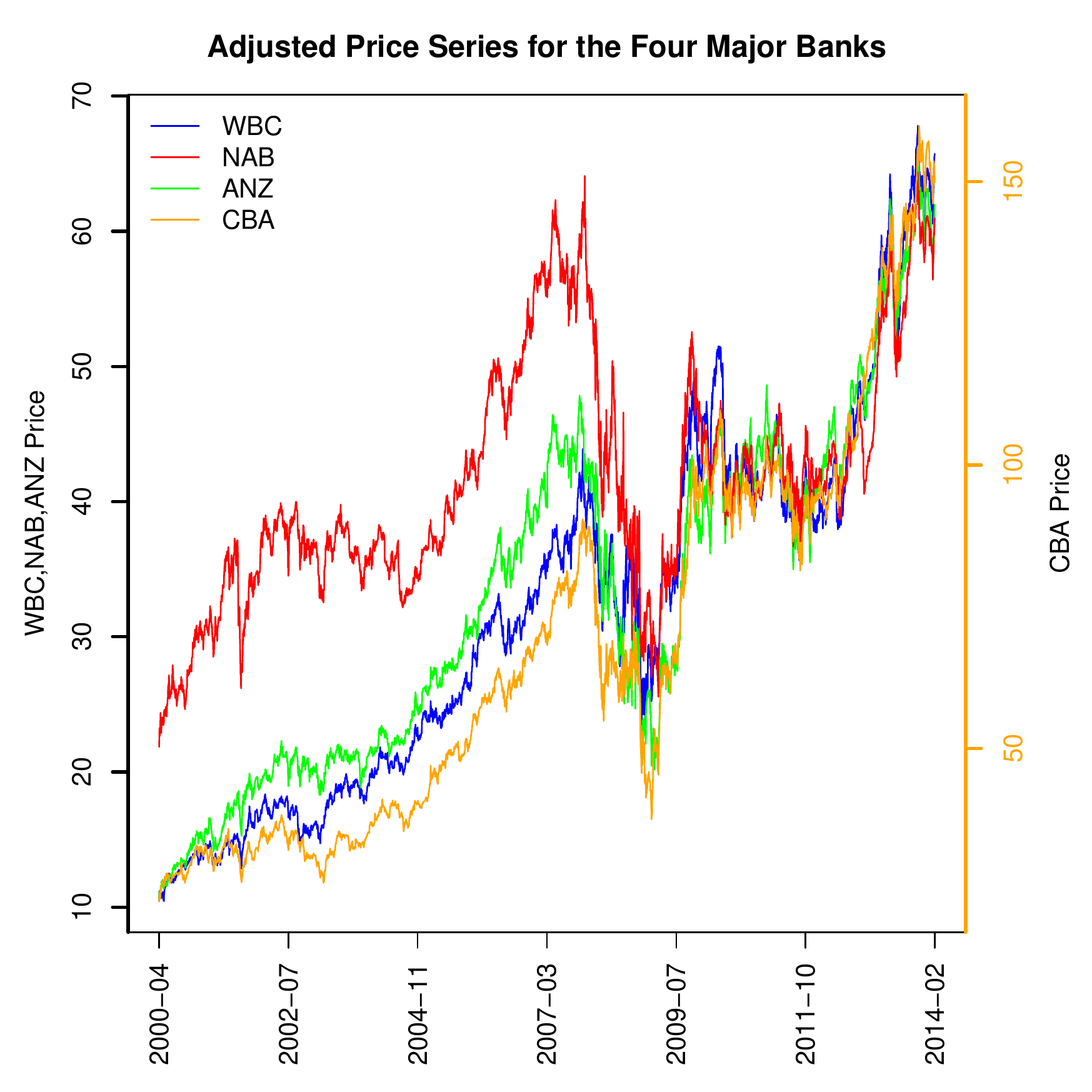}}
\end{figure}

Even if, say, a pair of highly correlated stocks are 
identified there may be considerations
other than diversification which mean that an investor may hold
both. 
Figures (\ref{fig:CFXBHP}), (\ref{fig:MGRSGP}), and (\ref{fig:fourbigbanks})
show that high correlations in price movements do not necessarily indicate
that the returns over the longer term will be similar. Thus depending on
how much short-fall risk an investor is willing to take on, he or she
may decide to hold more than one of the stocks identified in each of
these groups.

We applied the PCA to the full sample period to illustrate the value of
the method. In practical applications
a fund manager would apply the PCA on a rolling window basis to a much shorter
period of data. However, our results show that highly correlated stocks
tend to remain highly correlated over long periods of times. Nevertheless,
these correlations may break down as is evident in Figure (\ref{fig:CFXBHP}).
It is clear that since about 2011 the strong correlation between
them evident in the first 11 years of the sample no longer held.

While the method presented here was applied to a stock market, it is
general and can be applied to any set of assets for which a correlation
matrix can be generated.

\bibliography{Diversification}

\begin{thebibliography}{}

\bibitem[\protect\citeauthoryear{Billio, Getmansky, Lo, and Pelizzon}{Billio
  et~al.}{2012}]{Billio2012}
Billio, M., M.~Getmansky, A.~W. Lo, and L.~Pelizzon (2012).
\newblock {Econometric measures of connectedness and systemic risk in the
  finance and insurance sectors}.
\newblock {\em Journal of Financial Economics\/}~{\em 104}, 535--559.

\bibitem[\protect\citeauthoryear{Driesson, Melenberg, and Nijman}{Driesson
  et~al.}{2003}]{Driesson2003}
Driesson, J., B.~Melenberg, and T.~Nijman (2003).
\newblock {Common factors in international bond returns}.
\newblock {\em Journal of International Money and Finance\/}~{\em 22},
  629--656.

\bibitem[\protect\citeauthoryear{Jolliffe}{Jolliffe}{1986}]{Jolliffe1986}
Jolliffe, I.~T. (1986).
\newblock {\em {Principal Component Analysis}}.
\newblock New York: Springer.

\bibitem[\protect\citeauthoryear{Kim and Jeong}{Kim and Jeong}{2005}]{kim2005}
Kim, D.-H. and H.~Jeong (2005).
\newblock {Systematic analysis of group identification in stocks markets}.
\newblock {\em Physical Review E\/}~{\em 72}, 046133.

\bibitem[\protect\citeauthoryear{Kritzman, Li, Page, and Rigobon}{Kritzman
  et~al.}{2011}]{Kritzman2011}
Kritzman, M., Y.~Li, S.~Page, and R.~Rigobon (2011).
\newblock {Principal Components as a measure of systemic risk}.
\newblock {\em Journal of Portfolio Management\/}~{\em 37}, 112--126.

\bibitem[\protect\citeauthoryear{P\'{e}rignon, Smith, and Villa}{P\'{e}rignon
  et~al.}{2007}]{C2007}
P\'{e}rignon, C., D.~R. Smith, and C.~Villa (2007).
\newblock {Why common factors in international bond returns are not so common}.
\newblock {\em Journal of International Money and Finance\/}~{\em 26},
  284--304.

\bibitem[\protect\citeauthoryear{{R Core Team}}{{R Core Team}}{2014}]{R}
{R Core Team} (2014).
\newblock {\em R: A Language and Environment for Statistical Computing}.
\newblock Vienna, Austria: R Foundation for Statistical Computing.

\bibitem[\protect\citeauthoryear{Zheng, Podobnik, Feng, and Li}{Zheng
  et~al.}{2012}]{Zheng2012}
Zheng, Z., B.~Podobnik, L.~Feng, and B.~Li (2012).
\newblock {Changes in cross-correlations as an indicator for systemic risk}.
\newblock {\em Scientific Reports\/}~{\em 2}, 888.

\end{thebibliography}

\newpage

\begin{table}
\caption{Eigenvalues and variances explained by the last six principal components.}
\centering
\begin{tabular}{c |c c }
\toprule
  & \textbf{Eigenvalue} & \textbf{Variance explained(\%)} \\\hline
\textbf{PC151} & 0.40 & 0.25\% \\
\textbf{PC152 }& 0.38 & 0.24\% \\
\textbf{PC153 }& 0.32 & 0.21\% \\
\textbf{PC154} & 0.30 & 0.19\% \\
\textbf{PC155} & 0.29 & 0.18\% \\
\textbf{PC156 }& 0.24 & 0.16\% \\
\bottomrule
\end{tabular}
\label{table:eigenlast}
\end{table}

\newpage

\newpage

\newpage

\newpage

\newpage

\newpage

\newpage

\newpage

\end{document}